# Estimation of mass thickness response of embedded aggregated silica nanospheres from high angle annular dark-field scanning transmission electron micrographs

Matias Nordin[a], Christoffer Abrahamsson[a], Charlotte Hamngren Blomqvist[b], Magnus Röding[d], Eva Olsson[b], Magnus Nydén[c] and Mats Rudemo[d].

[a] Applied Surface Chemistry, Department of Chemical and Biological Engineering, Chalmers University of Technology, 412 96 Gothenburg, Sweden.

b Department of Applied Physics, Chalmers University of Technology, 412 96 Gothenburg, Sweden

c Ian Wark Research Institute, University of South Australia, SA 5095 Adelaide, Australia

d Department of Mathematical Sciences, Chalmers University of Technology and Gothenburg University, 41296 Gothenburg Sweden

## SUMMARY

In this study we investigate the functional behavior of the intensity in high-angle annular dark field (HAADF) scanning transmission electron micrograph (STEM) images. The model material is a silica particle (20 nm) gel at 5 wt%. By assuming that the intensity response is monotonically increasing with increasing mass thickness of silica, an estimate of the functional form is calculated using a maximum likelihood approach. We conclude that a linear functional form of the intensity provides a fair estimate but that a power function is significantly better for estimating the amount of silica in the z-direction.

The work adds to the development of quantifying material properties from electron micrographs, especially in the field of tomography methods and three-dimensional quantitative structural characterization from a STEM micrograph. It also provides means for direct three-dimensional quantitative structural characterization from a STEM micrograph.

## INTRODUCTION

Material characterization using transmission electron microscopy is a very active field. In particular, tomography methods have shown to be powerful (Midgley & Dunin-Borkowski (2009)) for direct access to the three-dimensional structure and subsequent morphological analysis (Gommes, Friedrich, Jongh & Jong (2010)). Of particular importance has been the development of the scanning transmission electron microscope (STEM) high angle annular dark field (HAADF), where the image intensity has an approximately monotonic relationship to the mass-thickness of the specimen (Hawkes (2005)) provided that the material studied is amorphous and that the collected electrons are incoherently scattered, which is known as the projection requirement (Midgley & Weyland (2003). The development of electron tomography methods has stimulated an increasing number of studies of controlled fabrication (Chiappini et al. (2010)), as well as characterization and reconstruction of materials in three dimensions (see e.g. Biermans, Molina, Batenburg, Bals & van Tendeloo (2010), Saghi et al. (2011), or the study by Xin, Ercius, Hughes, Engstrom & Muller (2010)). Material reconstruction by electron microscopy micrographs has its roots in the 1980's, where the main challenge was to prepare sufficiently thin samples in order to approximate the slice to a representative two-dimensional cut through the material (see e.g Weitz & Oliviera (1984)). It is worth emphasizing the possibilities of material characterization by direct thickness contrast imaging. However, to access this possibility, a functional form of the intensity response is needed first. In this study, we show that when the projection requirement holds, it is possible to directly estimate the intensity vs. mass-thickness function from the micrographs using maximum likelihood. We demonstrate this approach using aggregates of nanometer sized silica particles where they have aggregated to form a (very stable) particle gel.

## MATERIALS AND METHODS

The silica sol (BINDZIL 40/130) consisting of a 40 wt% aqueous dispersion of monodisperse silica spheres of diameter around 20 nm was kindly provided by EKA Akzo Nobel, Sweden. The sol pH was originally in the range of 9.10-9.20 and was adjusted to 7.8 by ion exchange (Dowex Marathon C) followed by suction filtration. The filtered sol, sodium chloride solution and deionized water was mixed, vortexed and left to gel for 14 days. Gel cubes of 1x1x1 mm were embedded in LR White resin (TAAB laboratories, England). The cubes were taken from the inner volume of the gel sample, discarding any surface areas. Prior to embedding, dehydration of the sample was performed in a graded ethanol series up to 99.5%. Ultra thin sections of approximately 90 nm (estimated from the colour of the reflectance of the section) were sliced using an ultramicrotome (Powertome XL, RMC products, Boeckeler Instruments Inc, Tucson, Arizona). The sections were placed on 200 mesh carbon support film Cu-grids and imaged in HAADF STEM mode with a Tecnai G2 (FEI Company, Eindhoven, Netherlands) using an accelerating voltage of 200 kV and with a camera length of 300 mm giving a HAADF detector inner radius of 22 mrad.

**STATISTICAL ANALYSIS**

Let us describe the sample by coordinates $(x, y, z)$ so that $\hat{z}$ describes penetration axis and $z_{max}$ indicates the thickness of the slice (in this study set to 90 nm). The mass thickness of the silica at $(x, y)$ can be written as a fraction of the total sample thickness

$$\alpha(x, y) = \frac{\int_0^{z_{max}} \rho_S(x, y, z) dz}{z_{max}} \quad (1)$$

where $\rho_S(x, y, z)$ denotes the silica density (depending on whether there is a silica particle at the point $(x, y, z)$ or not). Using this, we can write an estimate of the intensity in the direction normal to the plane of the sample (i.e. the z-direction) as

$$I(x, y) = \alpha S(\alpha) + (1 - \alpha) P(1 - \alpha) \quad (2)$$

where $S(\alpha)$ and $P(1 - \alpha)$ denote the intensity response with respect to the mass thickness of silica and the embedding polymer, respectively. The goal of this study is to estimate the functional form of the intensity response $I(x, y)$ with respect to $\alpha$ and our ansatz is $\alpha S(\alpha) + (1 - \alpha) P(1 - \alpha) = \alpha^\beta$ where $\beta$ is found by maximizing the log-likelihood function.

When the projection requirement holds, the intensity increases monotonically with increasing mass thickness, and the expression for the intensity can be expanded in powers of $\alpha$. In particular, where the response from the embedding polymer is weak, an expansion of the silica thickness is sufficient for good estimates of the response. Taking into account a base level intensity $b$ and a

function describing the random noise $e_{x,y}$, a simple model describing the observed intensity is

$$I(x,y) = b + cg(\alpha(x,y)) + e_{x,y} \quad (3)$$

where $c$ is a constant and $g$ is below specified as a power function. We assume that the random noise $e_{x,y}$ is well described by a normal distribution $N(0, \sigma^2)$ and that noise from different pixels $(x, y)$ are independent. Although this approach neglects diffraction effects (see e.g. Midgley & Weyland (2003)) we have found that under adequate conditions this approximates the intensity response in the system studied here well. By rewriting the fraction of silica in z-direction $\alpha$, as a function of particle center positions $x_i$ and defining a combined parameter-state vector $\theta = (N, \beta, \sigma, b, c, x_0, x_1, ..., x_N)$, the log-likelihood function $l_\chi(\theta)$ for the image data becomes

$$l(\theta) = -|M|\log(\sqrt{2\pi}\sigma) - \frac{1}{2\sigma^2} \sum_{x,y \in M} [I(x,y) - b - cg(\alpha(x,y))]^2 \quad (4)$$

that can be maximized using standard methods (e.g. simulated annealing).

**RESULTS AND DISCUSSION**

A representative micrograph is presented in the top left corner of figure 1 (top left image) taken of 5.5 wt% nano silica prepared as described above. By directly measuring the intensity in the micrograph it was noted that four rather separate values were obtained. These were attributed to the noise alone, or to one, two or three silica particles projected together, which was used as a starting value in the maximization of equation 4, where also the exponent $\beta$ was initially set to one. By this maximization an estimate of the power function intensity response $I(\alpha) = b + c\alpha^\beta$ was obtained with $b = 0.08$, $\beta = 0.69$ and the constant $c = 0.64$. This function is shown in figure 2 (blue line) where the error bars show the estimated standard deviation of the noise $\sigma$. Also shown (black line) is an alternative model, where the intensity response was estimated using a linear function (i.e. $\beta$ was kept at 1 and $c$ was estimated). The two models with a power function response and linear response are nested and can thus be tested with an approximate chi-square test from the log-likelihood fits. The test shows that with overwhelming significance (p-value <<0.0001) the power function gives a better fit. The top right micrograph in figure 1 shows the resulting re-generated micrograph. In the bottom left part of figure 1 the absolute difference between the experimentally obtained and the re-generated micrographs is shown. In figure 3 a histogram of the intensity of the micrograph (blue line), the re-generated image (black line) and the estimated noise (red). The background pixel intensity is added (no negative pixel values), which is why the estimated noise is shifted from origo. The re-generated micrograph (blue line) underestimates the intensity significantly in the range between 0.1 and 0.2. From investigating the estimated intensity response (figure 2) it is hypothesised that this range corresponds to the edges of one particle. In figure 1 the bottom right image shows a blow-up of the top left corner of the residual image, and indeed the errors seem to be located at the edges of the particles, which may partly be due to variation in particle size. Figure 2 also shows that the assumption of normal observation errors in equation (3) is reasonably adequate.

**CONCLUSIONS**

We have shown that it is possible to retrieve the functional form of the intensity response dependence of mass thickness by direct analysis of the micrographs. For this procedure to work, a well-defined material such as the mono-disperse and mono-phase particles or, as in this case, aggregates of particles must be used.

By knowing the mass thickness-intensity function, a three dimensional estimate of the studied sample can be made from the micrograph. The only assumption needed regarding the material structure is that all particles have aggregated (and that specimen surface effects from the thin film TEM sample preparation resulting in partial aggregates are avoided).

## ACKNOWLEDGEMENTS

The authors would like to thank Stefan Gustafsson (Chalmers University of Technology) for fruitful discussions and valuable comments. The work was funded by the Swedish Science Council (VR project no. 2008-3895) and the Vinnova financed VINN Excellence Center SuMo Biomaterials.

## REFERENCES

Biermans, E., Molina, L., Batenburg, K.J., Bals, S. & van Tendeloo, G. (2010) Measuring porosity at the nanoscale by quantitative electron tomography, Nano Letters 10, 5014-5019

Chiappini, C. et al. (2010) Tailored porous silicon microparticles: fabrication and properties, Chemphyschem 11, 1029-1035

Gommes, C. J., Friedrich, H., de Jongh, P. E. & de Jong, K. P. (2010) 2-Point correlation function of nanostructured materials via the grey-tone correlation function of electron tomograms: A three-dimensional structural analysis of ordered mesoporous silica, Acta Material 58, 770-780

P.W. Hawkes, The electron microscope as a structure projector, in: J. Frank (Ed.), Electron Tomography: Three- dimensional Imaging with the Transmission Electron Microscope, second edition, Plenum Press, New York, London, 2005.

Midgley, P. A. & Weyland, M. (2003) 3D electron microscopy in the physical sciences: the development of Z-contrast and EFTEM tomography, Ultramicroscopy 96, 413-431

Midgley, P. & Dunin-Borkowski, R. (2009) Electron tomography and holography in materials science, Nature Materials 8, 271-280

Saghi, Z. et al. (2011) Three-Dimensional Morphology of Iron Oxide Nanoparticles with Reactive Concave Surfaces. A Compressed Sensing-Electron Tomography (CS-ET) Approach, Nano Letters 11, 4666-4673

Weitz, D. A. & Oliviera, M. (1984) Fractal Structures Formed by Kinetic Aggregation of Aqueous Gold Colloids, Phys. Rev. Lett. 52, 1433-1437

Xin, H. L., Ercius, P. K., Hughes, J. J., Engstrom, R. & Muller, D. A. (2010) Three-dimensional imaging of pore structures inside low-kappa dielectrics, Appl. Phys. Lett. 96, 223108

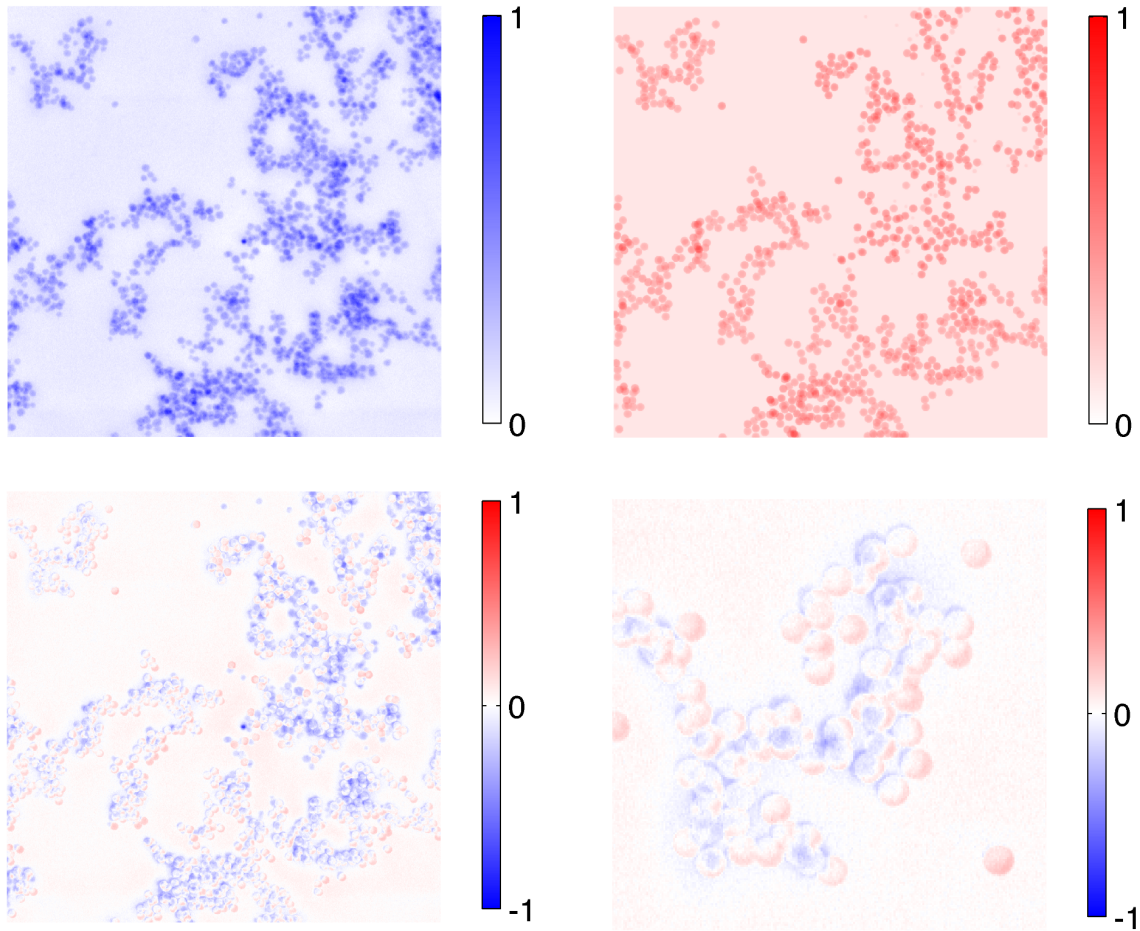

**Figur 1 Top left:** Micrograph of a 90 nm slice of 5 wt% aggregated nano silica obtained with HAADF-STEM. **Top right:** re-generated micrograph by maximizing the log-likelihood function (equation 4) using a power function as described. **Bottom left:** Residual image $R(x,y) - M(x,y)$, of the original micrograph $M$ and the re-generated one $R$. **Bottom right:** blow-up of the residual image showing the top left cluster.

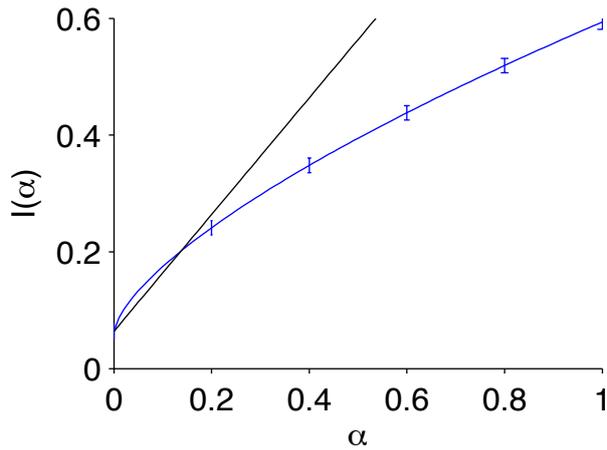

**Figur 2** The intensity response I vs. the amount of silica alpha as estimated by maximizing the log-likelihood function (equation 4). The micrographs are recorded from a specimen which was 90 nm in thickness and the silica spheres are 20 nm. One sphere corresponds to alpha = 0.22. The blue line shows an estimate using the $I(\alpha) = c\alpha^\beta$ model and the black line shows a linear model ($\beta = 1$ and $c$ is estimated). The error bars show the estimated standard deviation of the background noise.

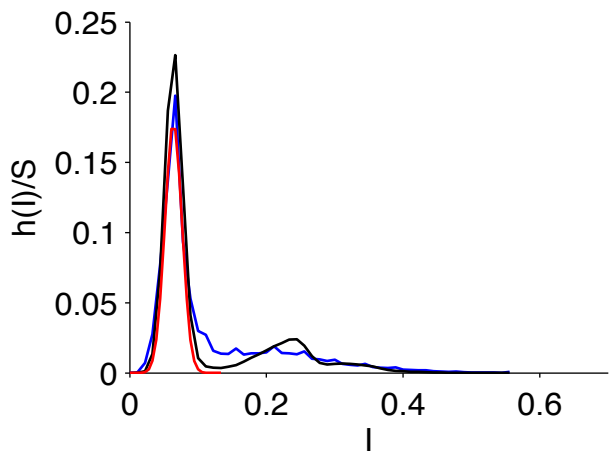

**Figur 3** Histogram of the pixel intensities for the micrograph (blue) and the generated micrograph (black). The estimated background noise is also shown (red). Note that the micrograph is more blurry at the edges of the silica spheres. This effect is not taken into account in the model which can explain why the generated data show an underestimate just between $I = 0.1$ and $0.2$